# Microwave power coupling in a surface wave excited plasma

Satyananda Kar*, Lukas Alberts and Hiroyuki Kousaka

*Department of Mechanical Science and Engineering, Nagoya University, Furo-cho, Chikusa-ku, Nagoya - 464 8603, Japan*

*babuliphy@gmail.com

**ABSTRACT**

In recent decades, different types of plasma sources have been used for various types of plasma processing, such as, etching and thin film deposition. The critical parameter for effective plasma processing is high plasma density. One type of high density plasma source is Microwave sheath-Voltage combination Plasma (MVP). In the present investigation, a better design of MVP source is reported, in which over-dense plasma is generated for low input microwave powers. The results indicate that the length of plasma column increases significantly with increase in input microwave power.

## I. INTRODUCTION

Recently, microwave excited plasma sources play a vital role in industry for surface modification and thin film deposition. Microwave excited plasmas are better known for generation of over-dense (greater than the cutoff plasma density, i.e., $7.6 \times 10^{10}$ cm$^{-3}$ for 2.45 GHz) plasmas in a large diameter geometry. Microwave excited plasma source is two types, (i) without external magnetic field, i.e., surface wave plasma (SWP) source and (ii) with external magnetic field, i.e., ECR plasma source.

SWP sources can be divided into dielectric-bounded type (surface waves propagate along the dielectric and plasma interfaces) and metal-bounded type, employing microwave sheath-voltage combination plasma (MVP), where surface waves propagate along the plasma and sheath interfaces.

In dielectric-bounded SWP sources, over-dense plasmas are sustained by the fields of electromagnetic waves travelling along plasma-dielectric (mostly quartz) surface boundaries [1-4]. Also peoples have used slot antennas for launching of microwave [4-8] for higher densities. In slot antenna configuration, plasma lengths are broadened with decrease in gas pressures [9]. In dielectric-bounded SWP sources, more than 500 W microwave power is injected.

In MVP sources, microwave plasmas are produced along a dielectric-free metal antenna and the surface waves propagate along the plasma-sheath interfaces. In 2005, Kousaka et. al. [10] at Nagoya University reported a new SWP source which can generate over-dense plasma along the surface of a negative biased metal rod, called metal-antenna surface wave excited plasma (MASWP). Later on, they named it MVP [11]. Their works [10-14] suggested that the ion sheath region between the negative biased metal rod and the plasmas acts as a passage for electromagnetic wave transmitting. Kousaka et. al. [13] have reported the pressure dependence in a MVP source, i.e., the plasma column length is increased with increase in gas pressure. They have observed that the length of plasma column approximately saturates after 33 Pa. Kousaka et. al. [14] also have reported that increasing microwave power is not effective to lengthen the plasma column. In those cases [13-14], the antenna was completely inside the vacuum chamber. Zhu et. al. [15] numerically have simulated a MVP source and suggested the surface plasmon polaritons can be excited in the ion sheath. In the above said MVP sources, the negative biased metal rod or antenna is inside the vacuum chamber and the microwave is coupling with plasma through a quartz plate. In this report, a better design of a MVP source is reported, where the input microwave power can couple easily with plasma and can generate over-dense plasma for low input microwave powers.

However, till now, the production mechanism of MVP (sustaining of plasma) is not clear. MVP is expected as one of promising method for ultra-high-speed coating of diamond like carbon (DLC) [16-20], and internal DLC coating of narrow tubes [21-23]. However, the physics behind the MVP sources, particularly the physics of plasma generation therein are not well understood, and thus design principle for better reactor and recipes for DLCs has not yet been established.

## II. EXPERIMENTAL SET-UP

The schematic diagram of the experimental setup is shown in Fig. 1. The experiment was performed in a grounded cylindrical chamber of stainless steel (SUS304, JIS) with an inner diameter of 146 mm and a length of 500 mm. Mainly, this schematic consists of 5 components: (1) a vacuum chamber (anode/ground), (2) a coaxial electromagnetic wave-guide (WX-39D) for the microwave power feed-in, (3) a central conductor or antenna (Titanium of 18 mm of diameter and 300 mm of length inside the vacuum chamber) which is set to a negative DC bias (cathode), (4) a dielectric window (quartz of dielectric constant $\epsilon_r \approx 3.7$ and thickness of 10 mm) which seals the vacuum chamber to the coaxial wave-guide and couples the electromagnetic wave, and a Langmuir probe (cylindrical probe of 8 mm length and 0.2 mm diameter) for measurement of plasma parameters. The chamber was evacuated by a rotary pump and the base pressure was about 0.5 Pa. The working gas was argon in the pressure range of 40 Pa to 130 Pa.

The important part in our experimental system is the antenna. The antenna was long enough, i.e., 300 mm length inside the vacuum chamber and was extended from vacuum chamber to the waveguide. The old design [13-14] was used a special quartz window interrupting the antenna (detrimental to the microwave power coupling) and needed supplementary DC connection feed through (2 port design). This new design feeds the DC and the microwave power from one port only. The waveguide part of the antenna was cooled by water during the experiments.

First, the cathode/antenna was set to a negative voltage (-100 V to -300 V), so a low density plasma can form with an ion sheath layer along the antenna. Then the microwave was fed into the coaxial wave-guide from the right side of the system. Now the input microwave power could propagate easily along the negative biased antenna and surface wave excited plasma generated along the antenna inside the chamber. The input microwave power was 30 W to 60 W. The surface wave excited plasma columns were photographed by a camera through a side viewport. A Langmuir probe (SmartProbe, Scientific Systems) was inserted in the axial direction (opposite direction of antenna and the gap between the Langmuir probe and antenna was 26 mm), equipped with an automatic linear drive to measure the axial variation of plasma properties.

## III. RESULTS AND DISCUSSIONS

Figure 2 shows the surface wave excited plasma columns along the antenna for various input microwave powers, $P_{in}$ (30 W, 40 W and 50 W). The generated plasma shows columnar structure surrounding the antenna surface. It is clearly seen that the length of plasma column increases with increase in $P_{in}$. Here the gas pressure was 130 Pa and the applied DC bias was -250 V with the discharge currents of 0.19 - 0.39 A for various $P_{in}$.

Figure 3 shows the axial distributions of ion density and electron temperature for various $P_{in}$. These measurements are taken from the left side of the system. In Fig. 3(a), the ion density values are more than $1 \times 10^{11}$ cm$^{-3}$, which is greater than the cutoff density $7.6 \times 10^{10}$ cm$^{-3}$ for 2.45 GHz microwave employed. Here, we can see that, the ion density distributions are flattened and higher, when $P_{in}$ increases. This means that the length of plasma column increases with increase in $P_{in}$. In Fig. 3(b), the distribution of electron temperature is more flatter with increasing $P_{in}$ and the electron temperatures are around 1 - 2.5 eV over the plasma column.

Figure 4 shows the axial ion density and electron temperature distributions for various gas pressures, where the applied DC bias was -250 V and $P_{in}$ was 30 W. Here the discharge currents were varied from 0.14 - 0.26 A for various gas pressures. Here the length of plasma column increases with increase in gas pressure, but plasma density decreases along the antenna for higher gas pressures. Figure 5 shows that length of plasma column and ion density increases with increase in applied DC bias to the antenna, where the gas pressure was 50 Pa and $P_{in}$ was 50 W. Here the discharge currents were varied from 0.29 - 0.39 A for various DC bias.

The increase of length of plasma column observed at higher gas pressures is due to the propagation of electromagnetic surface waves along the interfaces of over-dense plasma and the under-dense ion sheath. Here the sheath region plays an important role. In Fig. 4, it is clearly seen that plasma density decreases with increase in gas pressure. So the ion sheath thickness will be increased. Because, the Child-Langmuir theory indicates that the sheath thickness is proportional to the three-fourth power of sheath voltage (applied DC bias) and the inverse of the square root of plasma density [24], hence sheath thickness is expanded by both increasing sheath voltage and decreasing plasma density. So when sheath thickness increases, the surface waves smoothly propagate longer distance along the sheath-plasma interface. In Fig. 5, plasma column is lengthened for increasing applied DC bias. If applied DC bias is increased, the sheath thickness would be expanded, implies a longer plasma column.

In Fig. 3(a), it is seen that length of plasma column and plasma density increase with increase in $P_{in}$ for constant DC voltage and gas pressure. So the sheath thickness would be decreased. Here is a good coupling of input microwave power and plasma through the extended antenna and the microwave power absorbed by the plasma would be very less due to high plasma density. Microwave power can't penetrate into over-dense plasma. So the high input microwave power drives a longer plasma column.

After all, for these high gas pressures, ion sheaths are collisional, because ion-neutral collision mean free path (0.016 cm for 130 Pa) is less than the ion sheath thickness (0.165 cm for -250 V, electron temperature of 2 eV and plasma density of $1 \times 10^{17}$ m$^{-3}$). Also titanium atoms may be present inside the sheath due to sputtering. So in the collisional ion sheath, charge exchange collisions between the ions and atoms will change the kinetic energy and density of ions, i.e., will reduce the flux inside the sheath [25]. This charge exchange collision may be a factor for easy propagating of surface waves.

In dielectric-bounded SWP sources, the input microwave power is more than 500 W. But in the present MVP source, we can generate over-dense plasmas for relatively low input microwave powers, around 30 W, which is shown in our experiments. Also MVP sources are free from undesirable impurities, because surface waves propagate along the sheath-plasma interface, unlike the dielectric-bounded type.

## IV. SUMMARY

In summary, a new MVP source is reported, in which the input microwave power couples with plasma very efficiently through an extended antenna. So over-dense plasma (around $2 \times 10^{11}$ cm$^{-3}$) can be generated for low input microwave power in such type of MVP sources. Length of plasma columns directly proportional to input microwave power, gas pressure and applied DC voltage in such type of MVP sources. The present experimental results suggest that MVP source can be used for ultra-high-speed DLC coating and would be a suitable plasma processing device. It has the potential to replace the big vacuum chambers with batch processing of a lot of samples, to some part-2-part processing with low cycle time.

**Acknowledgement**
This work was supported partly by DAIKO Foundation RESEARCH FELLOWSHIP PROGRAM in FY2014 and a "Grant for Advanced Industrial Technology Development (No. 11B06004d)" in 2011 from the New Energy and Industrial Technology Development Organization (NEDO) of Japan.

**REFERENCES**
1. H. Kousaka and K. Ono, *Plasma Sources Sci. Technol.* **12**, 273 (2003).
2. M. Nagatsu, K. Naito, A. Ogino and S. Nanko, *Plasma Sources Sci. Technol.* **15**, 37 (2006).
3. S. Morita, M. Nagatsu, I. Ghanashev, N. Toyoda and H. Sugai, *Jpn. J. Appl. Phys.* **37**, L468 (1998).
4. H. Sugai, I. Ghanashev and M. Nagatsu, *Plasma Sources Sci. Technol.* **7**, 192 (1998).
5. I. Odrobina, J. Kudela and M. Kando, *Plasma Sources Sci. Technol.* **7**, 238 (1998).
6. F. Werner, D. Korzec and J. Engemann, *Plasma Sources Sci. Technol.* **3**, 473 (1994).
7. Z. Q. Chen, m. H. Liu, P. Q. Zhou, W. Chen, C. H. Lan and X. W. Hu, *Plasma Sci. Technol.* **10**, 655 (2008).
8. X. Xu, F. Liu, Q. H. Zhou, B. Liang, Y. Z. Liang and R. Q. Liang, *Appl. Phys. Lett.* **92**, 011501 (2008).

9. I. Ghanashev, M. Nagatsu and H. Sugai, *Jpn. J. Appl. Phys.* **36**, 337 (1997).
10. H. Kousaka, J. Q. Xu and N. Umehara, *Jpn. J. Appl. Phys.* **44**, L1052 (2005).
11. H. Kousaka, H. Iida and N. Umehara, *J. Vac. Soc. Jpn.* **49**, 183 (2006). [in japanese]
12. H. Kousaka and N. Umehara, *Vacuum* **80**, 806 (2006).
13. H. Kousaka, J. Q. Xu and N. Umehara, *Vacuum* **80**, 1154 (2006).
14. H. Kousaka and N. Umehara, *Trans. Materials Research Soc. Jpn* **31**, 487 (2006).
15. L. J. Zhu, Z. Q. Chen, Z. X. Yin, G. D. Wang, G. Q. Xia, Y. L. Hu, X. L. Zheng, M. R. Zhou, M. Chen and M. H. Liu, *Chin. Phys. Lett*. **31**, 035203 (2014).
16. X. Deng, Y. Takaoka, H. Kousaka and N. Umehara, *Surf. Coat. Technol*. **238**, 80 (2014).
17. X. Deng, H. Kousaka, T. Tokorayama and N. Umehara, *Tribology Online* **8**, 257 (2013).
18. H. Kousaka, Y. Takaoka and N. Umehara, *Procedia Engineering* **68**, 544–549 (2013).
19. H. Kousaka, T. Okamoto and N. Umehara, *IEEE Transactions on Plasma Science* **41**, 1830 (2013).
20. Y. Takaoka, H. Kousaka and N. Umehara, *Extended Abstract of 13$^{th}$Plasma Surface Engineering*, OR1807 (4pages) (2012).
21. H. Kousaka, K. Mori, N. Umehara, N. Tamura and T. Shindo, *Surf. Coat. Technol*. **229**, 65 (2013).
22. R. Matsui, K. Mori, H. Kousaka and N. Umehara, *Diamond and Related Materials* **31**, 72 (2013).
23. R. Matsui, H. Kousaka and N. Umehara, *Jpn. J. Appl. Phys*. **52**, 11NA01 (2013).
24. M. A. Lieberman and A. J. Litchenberg, Principles of Plasma Discharges and Materials Processing, Wiley, New York, 1994.
25. T. Ibehej and R. Hrach, *WDS'13 Proceedings of Contrbuted Papers, Part II*, 54-59 (2013).

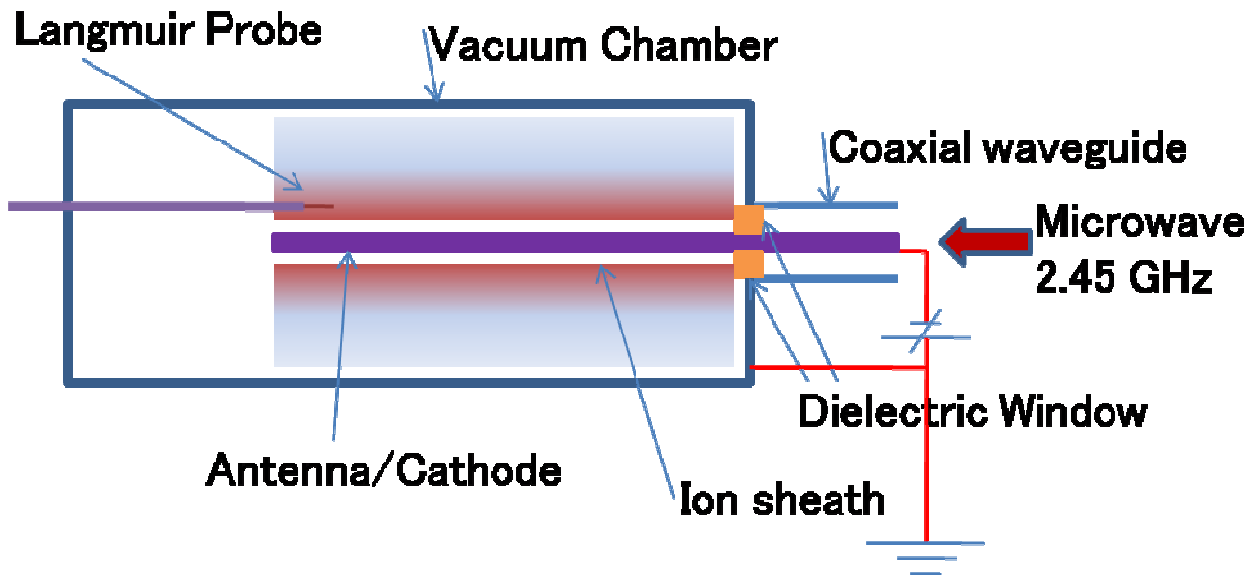

**Fig. 1.** Schematic of the experimental setup. Length of the vacuum chamber is 50 cm and inner diameter is 14.6 cm.

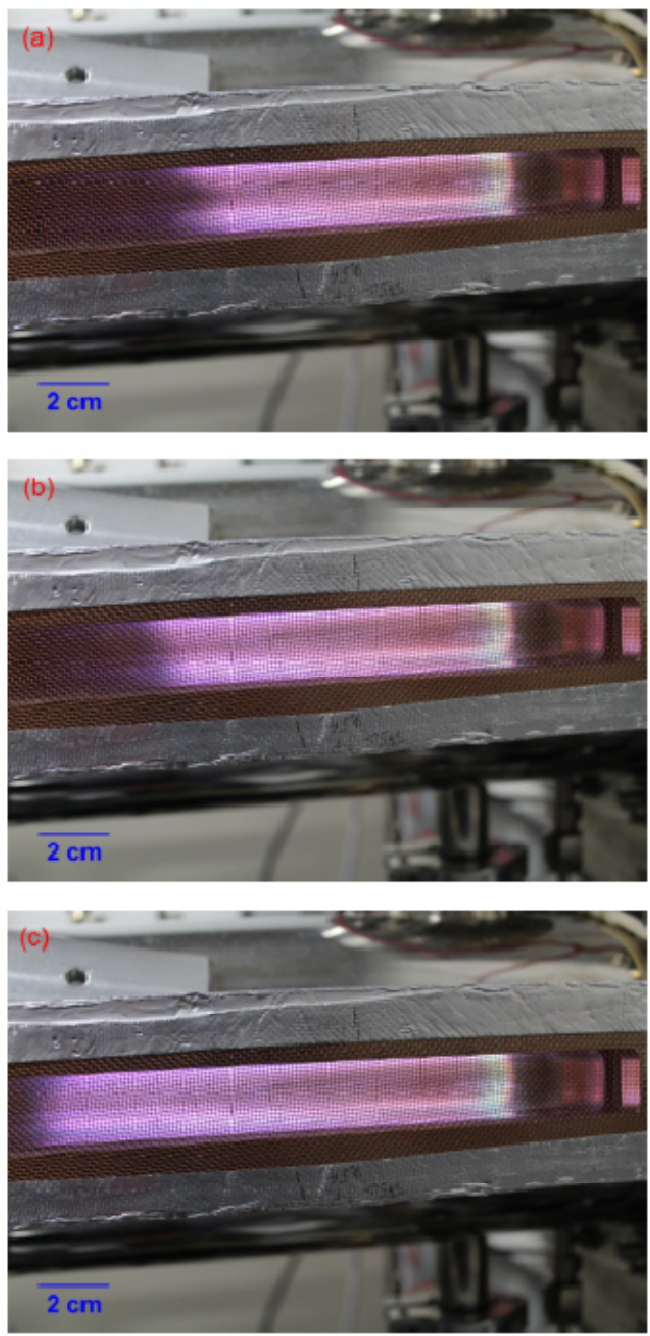

**Fig. 2.** Photographs of the plasma columns. (a) for 30 W, (b) for 40 W and (c) for 50 W. The applied DC bias (-250 V) and the gas pressure (130 Pa) were fixed.

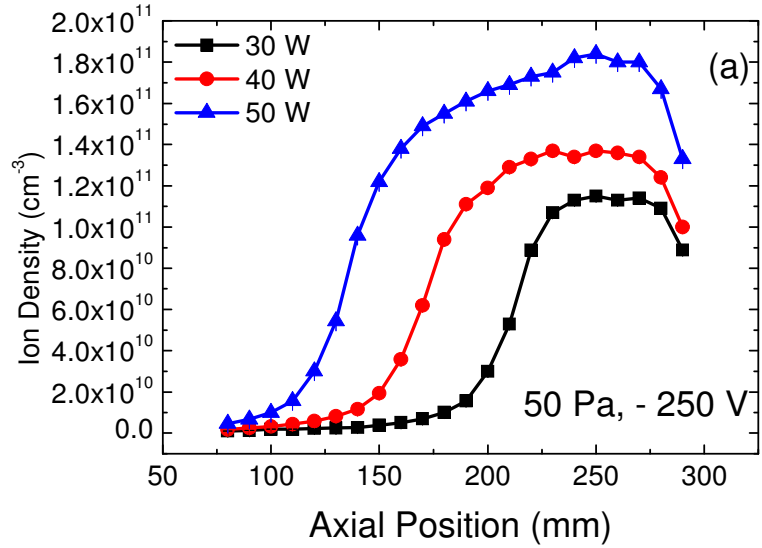

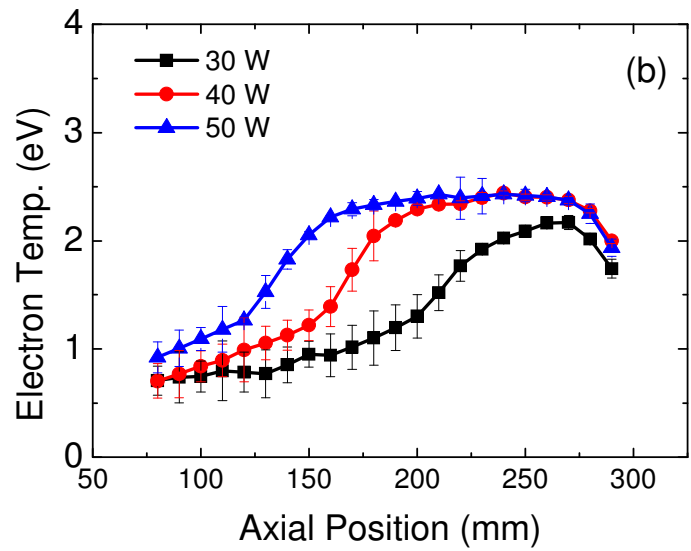

**Fig. 3** Axial distribution of (a) Ion density and (b) electron temperature. Here the applied DC bias (-250 V) and gas pressure (50 Pa) were fixed. The discharge currents were varied from 0.14 - 0.35 A for various $P_{in}$.

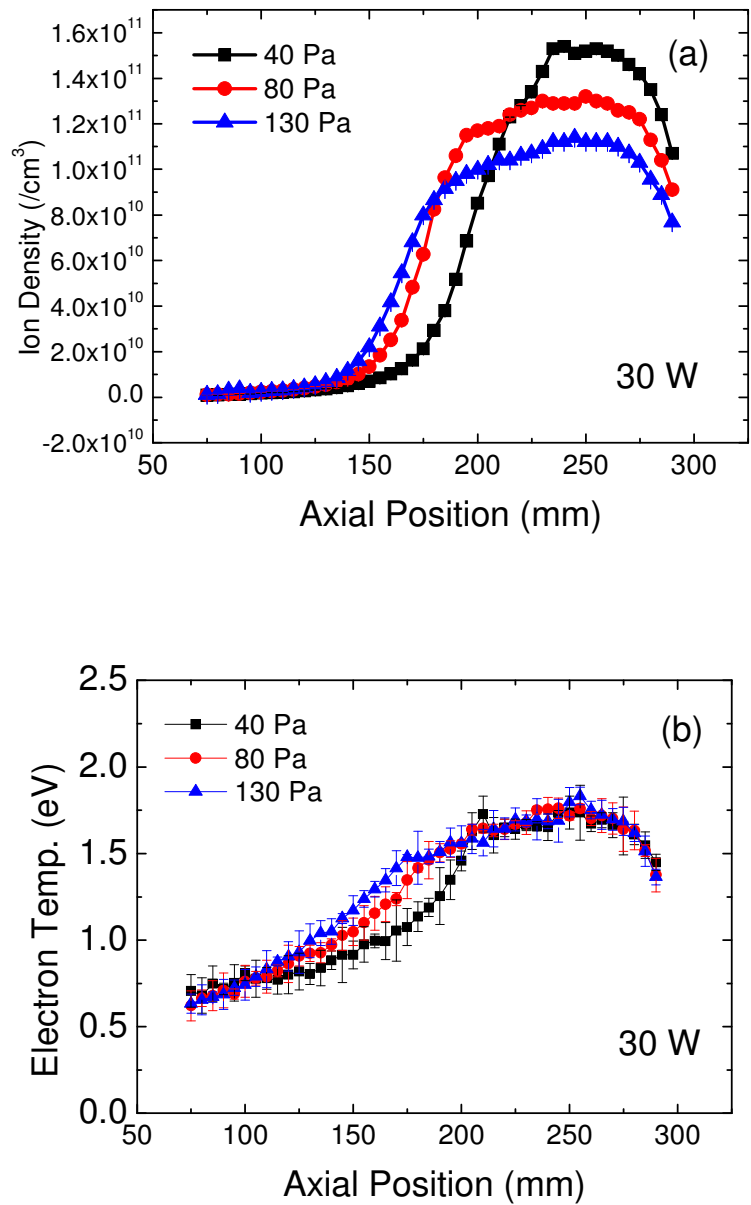

**Fig. 4.** Axial distribution of (a) ion density and (b) electron temperature for various gas pressures. Here the applied DC bias (-250 V) and $P_{in}$ (30 W) were fixed.

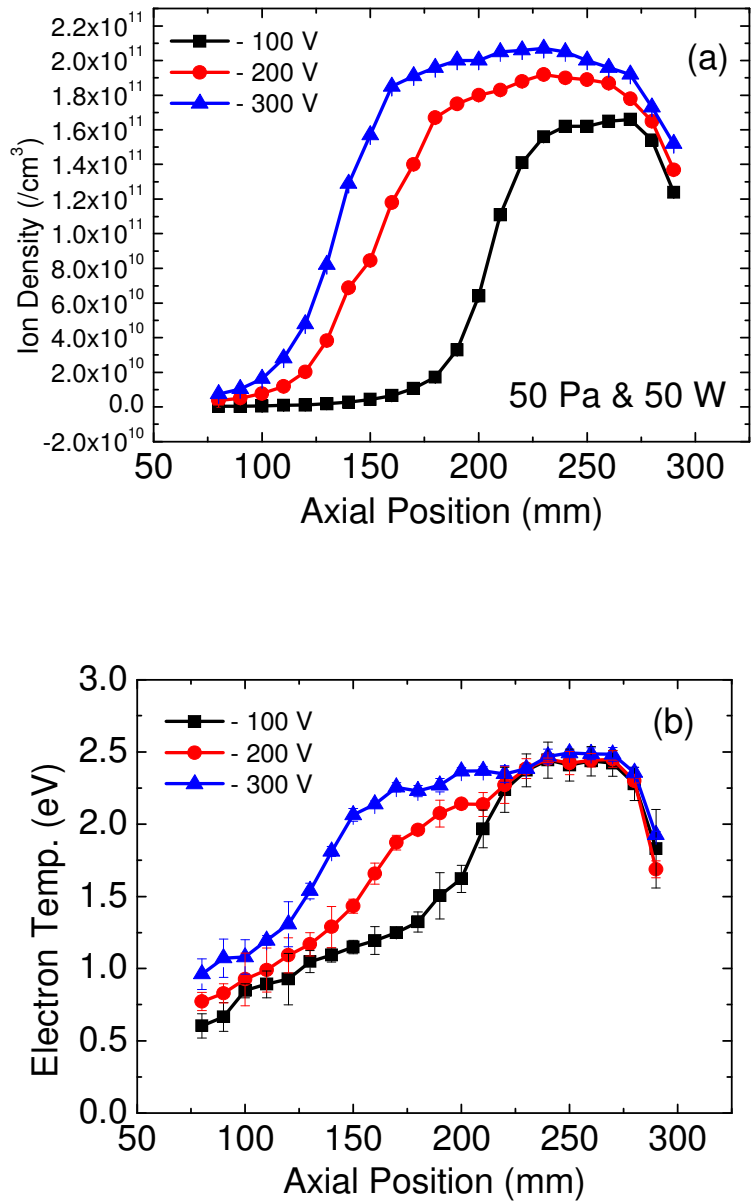

**Fig. 5.** Axial distribution of (a) ion density and (b) electron temperature for various DC voltages. Here the input microwave power (50 W) and the gas pressure (50 Pa) were fixed.